\documentclass[12pt, executivepaper]{article}
\usepackage{times}
\usepackage[utf8]{inputenc}
\usepackage[T1]{fontenc}
\usepackage{amsmath, xfrac}

\usepackage{color}
\usepackage{amsfonts}
\usepackage{graphicx}
\usepackage{setspace}
\usepackage{enumitem}
\usepackage{geometry}
\usepackage{dcolumn, braket, graphicx, xcolor, bm, lipsum, cases, epigraph}
 \usepackage{mathptmx}
\usepackage{hyperref}
\hypersetup{colorlinks= true, allcolors=blue}
\usepackage{natbib}

\setcitestyle{authoryear,open={(},close={)}}

\title{\textbf{Two Wrongs Make One Right}}

\author{Michele Pizzochero\thanks{E-mail: mp2834@bath.ac.uk} \\   \emph{\small{Department of Physics, University of Bath}},  \emph{\small{Bath BA2 7AY, United Kingdom}}   \\  \emph{\small{School of Engineering and Applied Sciences, Harvard University}},   \emph{\small{Cambridge, MA 02138, United States} } }

\date{}
\begin{document}

\maketitle 

\begin{abstract}
An influential argument for scientific realism posits that, if scientific theories were not true, their empirical success would be a coincidence. Here, I show that the false Drude's theory succeeds in explaining the Wiedemann-Franz law by coincidence---a fortuitous compensation of errors.  \\ 


\end{abstract}
\vskip -0.2cm
%\bigskip
%\bigskip
%\bigskip
%\bigskip
\setlength{\epigraphwidth}{0.615\textwidth}
\epigraph{If you had committed only one error, you would not have come at a true Solution of the Problem. 
But by virtue of a twofold mistake you arrive, though not at Science, yet at Truth.}{Bishop George Berkeley, The Analyst (1734)}

\tableofcontents
\bigskip
\bigskip
%\subsection*{Acknowledgments} 
%This work is dedicated to the students of my Condensed Matter Physics module at the University of Bath---with the hope that it will spark their curiosity to go beyond the `shut up and calculate' approach that too often shapes physics education.

\clearpage

\section{Truth, Coincidences, and Scientific Success} 
\label{Sec-1} 
That science is empirically successful is seldom disputed. But how should we explain this success? This question has fueled a century-long debate among scientists and philosophers, dividing them into two camps. Realists, on the one hand, hold that the empirical success of scientific theories originates from their ability to reveal objective truths about reality, particularly concerning the existence and behavior of unobservable entities. Antirealists, on the other hand, deny this view.\footnote{Providing a comprehensive overview of the scientific realism debate would be impractical. Therefore, the discussion here is restricted to some key ideas that have emerged in the debate. For a detailed discussion of scientific realism, see, e.g., \cite{Psillos:1999} and \cite{Lyons2025}.}

\paragraph{\emph{The realist explanation of scientific success.}} The  `ultimate argument' for scientific realism traces its origin to \cite{Smart1963, Smart1968} and \cite{Putnam1975, Putnam1978}, and later received a robust formulation by, e.g., \cite{Musgrave1988} and \cite{Leplin1997}. Were it not for scientific realism, \cite{Smart1963} argues, the success of scientific theories should be regarded as an unlikely stroke of luck:  ``If we interpret a theory in a realist way, then we have no need for such a cosmic coincidence: it is not surprising that galvanometers and cloud chambers behave in the sort of way they do, for if there really are electrons, etc., this is just what we should expect. A lot of surprising facts no longer seem surprising'' (p.\ 39). Indeed,  \cite{Smart1968} continues,   ``If there were no such things'' as  ``electrons or whatever is postulated by the theory,'' then we  ``would have to suppose that there were innumerable lucky accidents in the behavior of things mentioned'' (p.\ 150).

In a seemingly similar vein, \cite{Putnam1975} claims that  ``the positive argument for realism is that it is the only philosophy that doesn’t make the success of science a miracle,'' asserting that  ``terms in a mature science typically refer''  and   ``theories accepted in a mature science are typically approximately true'' (p.\ 73). Otherwise, \cite{Putnam1978} insists, one  ``has to leave it without explanation'' that ``‘electron calculi' [...] correctly predict observable phenomena'' despite there being  ``no electrons [...].  If there are such things, then a natural explanation of the success of these theories is that they are partially true accounts of how they behave'' (pp.\ 18-19). 

Using the example of the electron, both Smart’s and Putnam’s arguments center on the idea that only two hypotheses can elucidate the success of science: either its theories are (approximately) true by correspondence---that is, they provide a faithful description of the unobservable aspects of external reality---or their success is nothing short of miraculous or coincidental. This conclusion has been framed by \cite{Musgrave1988} as an inference to the best explanation (also known as abduction), a non-deductive mode of reasoning that, in \cite{Harman1965}'s now-classic definition, moves  ``from the premise that a given hypothesis would provide a ‘better’ explanation for the evidence than would any other hypothesis, to the conclusion that the given hypothesis is true'' (p.\ 89). This inference is typically implemented through a comparative-eliminative process, which consists of ranking the existing candidate explanations and discarding the unlikely ones. As a result, the empirical success of scientific theories is better explained by their approximate truth than by competing explanations involving the occurrence of coincidences (according to Smart) or miracles (according to Putnam).\footnote{In addition to being framed as an inference to the best explanation, the argument has also been formulated in other ways, such as a Bayesian argument or a plausibility argument, as discussed by, e.g., \cite{Psillos:1999}.}

\paragraph{\emph{The antirealist explanations of scientific success.}}

The antirealist rejects the success-to-truth inference by advancing a two-pronged argumentative strategy that involves a \emph{pars destruens} and \emph{pars construens}, each spanning a wide spectrum of views. The antirealist \emph{pars destruens}  seeks to sever the connection between empirical success and approximate truth that resides at the core of scientific realism. The key argument against this connection is \cite{Laudan1981}'s pessimistic meta-induction, which highlights that history is rife with once empirically successful theories that are patently false according to our current science, with their central terms failing to refer. Since our current best theories are fundamentally similar to these past theories, we have inductive reasons to expect that they, too, will ultimately be proven false. Drawing from a long list of abandoned theories, \cite{Laudan1981} identifies a twofold pattern that divorces truth from success. On the one hand, ``a theory may be empirically successful even if it is not approximately true,''  that is,  the ``theory's success is no warrant for the claim that all or most of its central terms refer'' (p.\ 47), as exemplified by the phlogiston theory. On the other hand,  ``the fact that a theory's central terms refer does not entail that it will be successful'' (p.\ 47), as exemplified by the atomic theory in the eighteenth century.\footnote{Various arguments have been proposed against \cite{Laudan1981}'s view. For example, \cite{Psillos:1999} observes that many of the theories cited by \cite{Laudan1981} are not mature, as \cite{Putnam1975}'s original argument requires, thus weakening the strength of the inductive inference by reducing the pool of instances from which the inference is drawn. Additionally, certain theoretical claims that are responsible for the empirical success of abandoned theories have remained stable across theory change, leading to selective forms of scientific realism such as  \cite{Psillos:1999}'s deployment realism or \cite{Worrall1989}'s structural realism. I am here concerned mainly with standard scientific realism, although in Section \ref{Sec-5} I briefly discuss structural realism, likely the most promising strategy for the realist to address the challenge posed by Drude's predictive success. }

An additional challenge faced by scientific realism is the underdetermination of theories by evidence, occurring when two or more scientific theories entail indistinguishable observational consequences and yet offer conflicting descriptions of the unobservable reality. Because at least one of these theories ought to be false, despite being empirically equivalent to the true one, the realist is left uncertain on what exactly they are supposed to be realist about. \cite{Norton2008} outlines three mechanisms through which a given theory can be paired with an observationally indistinguishable rival: natural pairs, which arise during the natural development of scientific inquiry, such as the set of formulations of Newton's mechanics differing from distinct inertial states, as discussed by \cite{vanFraassen:1980}; cultured pairs, intentionally crafted for specific philosophical purposes, such as Poincaré-Reichenbach cases of multiple geometries; and artificial pairs, which involve quasi-scientific theories devised by philosophers for philosophical ends, such as \cite{Kukla1996}'s algorithms to generate infinite sets of  empirically equivalent theories.

The antirealist \emph{pars construens} proposes alternative accounts of scientific success, seeking to  show that truth is neither the only nor the best explanation. We can distinguish two classes of antirealist explanations, each of them encompassing various views: the `evolutionary' and the `empirical adequacy' explanations.  The evolutionary explanation aims to undermine the realist appeal to truth by focusing on the methodological structure of scientific practice. Scientific theories, on this view, emerge as a result of a rigorous process of scrutiny, testing, and refinement—``a winnowing process,'' in the words of \cite{Laudan1984}, ``that is arguably more robust and discriminating than any other method we possess for evaluating empirical claims about the world'' (p.\ 101)---by which theories that are not empirically adequate are systematically discarded. The success of science, as \cite{vanFraassen:1980} illustrates with an analogy, ``is not even surprising to the scientific (Darwinist) mind.'' Akin to natural selection in biological evolution, theories are ``born into a life of fierce competition, a jungle red in tooth and claw'' where only ``successful theories survive'' (p.\ 40).  On the evolutionary account, ``the best explanation for the success of science,'' as \cite{Wray2018} puts it,  ``is the fact that unsuccessful scientific theories have been abandoned'' (Chap.\ 1).\footnote{The evolutionary explanation of scientific success has been criticized by \cite{Leplin1997}, who notes that while it may explain \emph{how} we come to possess successful theories, it does not explain \emph{why} the successful theories we have are successful, i.e., what intrinsic property makes them successful: ``To explain why the theories \emph{that we select} are successful, it is appropriate to cite the stringency of our criteria for selection. But to explain why \emph{particular theories}, those we happen to select, are successful, we must cite properties of \emph{them} that have enabled them to satisfy our criteria'' (p.\ 9).}

Antirealists have offered various alternative metaphysical interpretations of the empirical success of scientific theories. For example, building on the earlier work of \cite{Fine1986}, \cite{Leplin1987} has introduced  `surrealism' (for ‘surrogate realism'), the view that a more metaphysically parsimonious explanation of scientific success can be achieved by deploying the  ‘as-if’ operator: instead of supposing that successful theories ``are true or partly true, or that the entities to which they purportedly refer exist,'' as the realist does, the antirealist can ``suppose only that the world behaves \emph{as if} this were the case'' (p.\ 520).\footnote{\cite{Leplin1997} criticizes surrealism by claiming that it is parasitic on scientific realism: ``If it is to explain success [...], surrealism must \emph{presuppose} that theoretical truth will be manifested in experience. But that presupposition is precisely the explanation realism gives to success [...]. Thus, surrealism is explanatory only in virtue of presupposing realist explanation'' (p.\ 27).} \cite{Stanford2000} has advanced a proposal that does not appeal to a relationship between theory and reality, but only between a pair of theories: by using the notion of `predictive similarity,' the success of a given theory ``is explained by the fact that its predictions are (sufficiently) close to those made by the true theoretical account of the relevant domain'' (p.\ 275), just like the predictions of the false Ptolemaic system closely approximate those of the true Copernican system.\footnote{\cite{Park2003} claims that \cite{Stanford2000}'s antirealist explanation is flawed because it fails to respect the ontological order between dependent and underlying properties, where the former should be explained in terms of the latter: ``The observational similarity [...] is the dependent property of the truth of the observational consequences''' and ``the truth of observational consequences is the underlying property of the observational similarity'' (p.\ 168). Additionally, \cite{Psillos2001} has raised several objections to \cite{Stanford2000}'s proposal, questioning its explanatory power and its adequacy as the final step in the chain of explanatory demands.} Other antirealists, for example \cite{vanFraassen:1980}, claim that a theory’s empirical adequacy \emph{is} the explanation of its success---end of story---rejecting any demand for further explanation.\footnote{As noted by \cite{Stanford2000}, as well as many others, ``Van Fraassen gives us no reason for ending our search for explanations with empirical adequacy, and no justification for refusing to answer the question  at just this point'' (p.\ 268).}

\paragraph{\emph{Miracles or coincidences?}}

As this brief overview suggests, a rarely acknowledged feature of the scientific realism debate is a point of agreement between realists and antirealists: neither camp is willing to entertain the idea that science succeeds by miracle or coincidence. The realist explicitly rejects this possibility: empirical success of science is best explained by the approximate truth of its theories, in that  it spares us from invoking coincidental or miraculous occurrences. The antirealist, while opposing the success-to-truth inference, is equally committed to avoiding any appeal to these explanations, instead articulating competing accounts of scientific success based on the methodology of scientific progress or alternative metaphysical readings. As a result, the hypothesis that a scientific theory can enjoy empirical success through miracles and coincidences remains quite unexplored, serving mainly as a provocative rhetorical device in the hands of the realist to dramatize the implausibility of antirealist positions.

More importantly, both realists and antirealists have treated Putnam's ``miracles'' and Smart's ``coincidences'' on an equal footing: absurd explanatory contenders. Although they both are used rhetorically to imply implausibility, their explanatory roles differ significantly in that, of course, miracles are \emph{not} coincidences.  Miracles do not constitute genuine explanations for a given fact. Rather, they are the absence, or refusal, of an explanation. This aligns with Putnam’s view that, without realism, scientific success would---interchangeably---be a ``miracle'' in \cite{Putnam1975} and ``without explanation'' in  \cite{Putnam1978}. Coincidences, on the other hand, are regarded by \cite{Smart1968} as ``lucky accidents'' and they contrast with miracles in that, as \cite{Smart1985} himself recognizes, ``we do not think of coincidences as inexplicable'' (p.\ 274).  Hence, ``it is illegitimate'' as  \cite{vanFraassen:1980} rightly notes, ``to equate being a lucky accident, or a coincidence, with having no explanation. It was by coincidence that I met my friend in the market---but I can explain why I was there, and he can explain why he came, so together we can explain how this meeting happened'' (p.\ 25). If the ultimate argument for scientific realism is construed as an inference to the best explanation, then it is questionable why, when relying on Putnam's version, it centers on juxtaposing a legitimate candidate explanation (i.e., the approximate truth of scientific theories) with a non-explanation (i.e., a miracle) to interpret scientific success. However, when relying on Smart's version, the argument is justified in including a coincidental occurrence among the potential explanations for scientific success, and then abductively concluding that it is a worse explanation than the rival one appealing to truth.

But could a scientific theory achieve empirical success by coincidence? The identification of a compelling case of a mature scientific theory that predicts phenomena at the observable level as a result of a ``lucky accident'' at the unobservable level would lend greater credibility to the hypothesis that empirical success can arise fortuitously. In what follows, I show that Drude's theory, in its account of the empirical regularity discovered by Wiedemann and Franz, realizes this  scenario. Echoing the example of electrons invoked in the arguments of \cite{Smart1963, Smart1968} and \cite{Putnam1978}, I will argue that Drude's theory does not achieve its success by delivering an approximately true description of how electrons behave, but rather through a coincidence: a surprising compensation of large errors. 

The remainder of this essay is structured as follows. Section \ref{Sec-2} presents the Wiedemann-Franz law, Drude's theory, and how the latter provides a quantitative prediction of the former. Section \ref{Sec-3} shows, through a comparison of Sommerfeld's theory, that the source of this empirical success lies in a fortuitous cancellation of errors. Section \ref{Sec-4} draws three key lessons for the realism-antirealism debate in light of this episode. Section \ref{Sec-5} outlines analogies and differences between Drude's theory and another historical case where coincidence seemed to have played an important role in the predictive success: Sommerfeld's prediction of the fine-structure formula. Finally, Section \ref{Sec-6} reiterates the importance of historical cases in the scientific realism debate and concludes this work.

\section{The Wiedemann-Franz Law and Drude's theory} 
\label{Sec-2} 
Metals comprise the majority of elements in the periodic table and are distinguished by their exceptionally high electrical and thermal conductivities. These two seemingly unrelated properties have been empirically linked through the regularity identified by Wiedemann and Franz. The effort to achieve a microscopic understanding of metallic behavior has played a central role in the development of condensed matter physics, with Drude's theory standing as a milestone in that pursuit.

\paragraph{\emph{The Wiedemann-Franz law.}} The electrical and thermal conduction of metals can be described through Ohm's and Fourier's laws, respectively, which were both introduced in the early eighteen hundreds. Ohm's law states that the electric current density $\vec{j}$ (i.e., the current flow per unit area) in a conductor is directly proportional to the applied electric field $\vec{\mathcal{E}}$, 
\begin{equation}
\vec{j} = \sigma \vec{\mathcal{E}}, 
\label{Ohm}
\end{equation}
where $\sigma$ is the electrical conductivity, an intrinsic, metal-dependent property that quantifies its ability to conduct electric current. Fourier's law states that the heat flux density $\vec{q}$ (i.e., the heat flow per unit area per unit time) is directly proportional to the negative temperature gradient $-\nabla{T}$,  
\begin{equation}
\vec{q} = -\kappa\nabla{T}, 
\label{Fourier}
\end{equation}
where $\kappa$ is the thermal conductivity, an intrinsic, metal-dependent property that quantifies its ability to conduct heat.

\begin{figure}[t!] 
    \centering
    \includegraphics[width=1\textwidth]{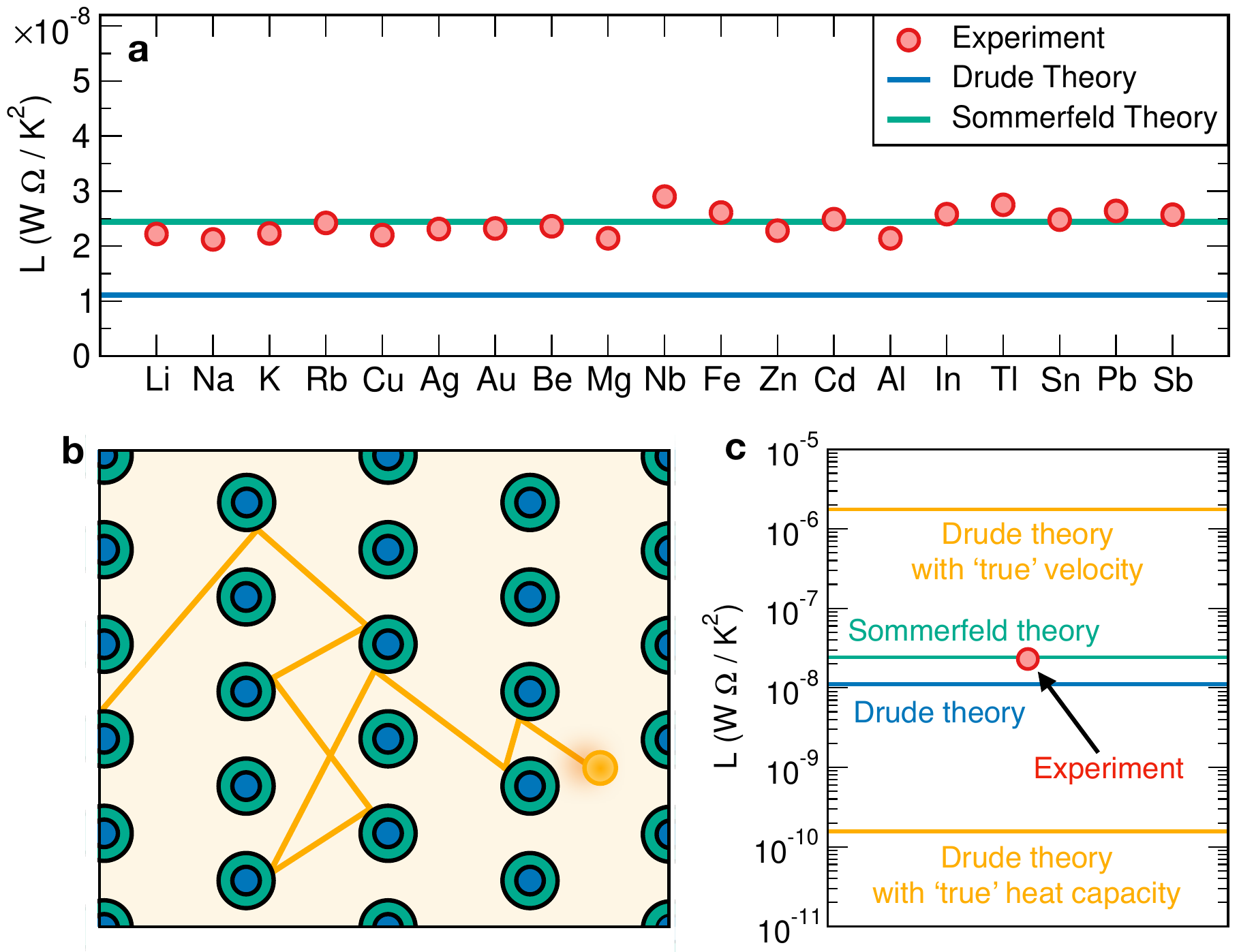} 
    \caption{\textbf{a}, Experimental values of the Lorenz number ($L$) for various metals measured at a temperature of 273 K (red dots), compared with the predictions of Drude's theory (blue line) and Sommerfeld's theory (green line); experimental values are from \cite{Ashcroft1976} (p.\ 21). \textbf{b}, Illustration of Drude's theory, where conduction electrons (small yellow circle) are pictured as classical particles that collide with immobile positively charged ions (large circles). \textbf{c}, Comparison of the Lorenz number ($L$) of silver (Ag) as determined from experimental measurements (red dot), Drude's theory (blue line), Sommerfeld's theory (green line), and the two fictitious versions of Drude's theory that include either the Sommerfeld electron velocity or Sommerfeld heat capacity (yellow lines); notice that $L$ is given on a logarithmic scale. }
    \label{Figure} 
\end{figure}

In the mid-eighteen hundreds, \cite{Franz1853} revealed an empirical regularity between the electrical and thermal conductivities: for a range  of different metals, these two quantities appeared in a fixed ratio for a given temperature. About two decades later, \cite{Lorenz1872} generalized these observations by demonstrating that  ``the ratio between the conductivity of a pure metal for heat and electricity is proportional to the temperature'' (p.\ 435). This ratio came to be known as the Wiedemann-Franz law,
\begin{equation}
\frac{\kappa}{\sigma} \cdot \frac{1}{T}  = L,
\label{Lorenz}
\end{equation}
where the proportionality constant $L$ is the Lorenz number, with a typical value of $L \approx 2.4 \times 10^{-8}$ W ${\Omega}$ / K$^2$. Although often treated as a universal constant---and indeed remarkably stable across a wide range of cases---the Lorenz number is weakly dependent on the specific metal and temperature. Figure \ref{Figure}(a) provides an overview of experimentally determined Lorenz numbers for a variety of representative metals at room temperature ($T = 273$ K), including alkaline metals (e.g., Li and K), alkaline earth metals (e.g., Be), transition metals (e.g., Fe and Ag), post-transition metals (e.g., Al and In), and metalloids (e.g., Sb). Among these, the highest Lorenz number is observed for Nb ($2.90 \times 10^{-8}$ W ${\Omega}$ / K$^2$) and the lowest for Na ($2.12 \times 10^{-8}$ W ${\Omega}$ / K$^2$). Similarly, the Lorenz number is quite insensitive to temperature. For example, increasing the temperature from 273 K to 373 K changes the Lorenz number of Be from $2.36 \times 10^{-8}$ W ${\Omega}$ / K$^2$ to $2.42 \times 10^{-8}$ W ${\Omega}$ / K$^2$ and that of In from $2.58 \times 10^{-8}$ W ${\Omega}$ / K$^2$ to $2.60 \times 10^{-8}$ W ${\Omega}$ / K$^2$, with variations across metals generally not exceeding $0.20 \times 10^{-8}$ W ${\Omega}$ / K$^2$, i.e., less than 10\%. For these reasons, the Lorenz number is widely considered independent of both the metal type and temperature.

\paragraph{\emph{Drude's theory.}} Right at the turn of the twentieth century,  \cite{Drude1900} introduced his `electron theory of metals.' As noted by \cite{Mott1936}, ``the greatest success of the theory was the explanation of the fact, discovered empirically by Wiedemann and Franz, that the ratio of the thermal to the electrical conductivity is approximately the same for all metals'' (p.\ xi). Some historical context is in order: when Drude worked out his theory of electrical and thermal conduction in metals, the electron had been discovered only three years earlier by Thomson, Rutherford's atomic model would not be proposed for more than a decade, and a comprehensive quantum theory of electrons in crystals lay far beyond the horizon. 

The central idea underlying Drude's theory is the extension of the kinetic theory of gases to metals, treated as a classical gas of electrons.\footnote{For a detailed presentation of Drude's theory, see, e.g., Chap.\ 1 of \cite{Ashcroft1976}, Chap.\ 4 of \cite{Alloul2011}, Chap.\ 3 of \cite{Simon2013}, Chap.\ 1 of \cite{Singleton2001}, and Chap.\ 5  of \cite{Hofmann2015}.} In the kinetic theory, gas molecules are depicted as identical, rigid spheres that travel in straight trajectories until they undergo instantaneous collisions with another molecule. While the simplest gases involve only one type of particle, a metal must host at least two, since the electrons carry a negative charge, yet the metal as a whole is electrically neutral. To account for this, Drude postulated the presence of a compensating positive charge, attributed to much heavier, immobile particles, which can be viewed as the ionic cores. A schematic illustration of Drude's theory is shown in Figure \ref{Figure}(b).
Drude pictured the behavior of electrons in metals according to three tenets. First, electrons form an ideal gas of classical particles. Second, electrons move freely in the metal, wandering until they suffer an instantaneous collision with an immobile ionic core that abruptly changes their velocity. Third, electrons experience no electromagnetic interaction either with ionic cores or with other electrons, reaching thermodynamic equilibrium solely through collisions.

On this basis, Drude derived a microscopic picture of Ohm's law, obtaining an expression for the current density $\vec{j}$ as a function of the electric field $\vec{\mathcal{E}}$,

\begin{equation}
\vec{j} = \frac{ne^2\tau}{m} \vec{\mathcal{E}}, 
\label{Drude-Ohm}
\end{equation}

\noindent
where $n \approx 10^{22}$ cm$^{-3}$  is the density of electrons in metals, $e = 1.60 \times 10^{-19}$ C is the elementary charge, $\tau \approx 10^{-14}$ s is a constant, mean free time between successive collisions, and $m = 9.10 \times 10^{-31}$ kg is the electron mass. By comparing Equation (\ref{Drude-Ohm}) with Equation (\ref{Ohm}), the electrical conductivity $\sigma$ can be identified as

\begin{equation}
\sigma = \frac{ne^2\tau}{m}.
\label{sigma}
\end{equation}

\noindent
Drude further regarded electrons as solely responsible for thermal conduction. Through the application of the kinetic theory of gases to electrons in metals, the following expression relating the heat flux density $\vec{q}$  to the temperature gradient $\nabla{T}$  can be obtained,

\begin{equation}
\vec{q} = -\frac{1}{3}v^2 \tau c_V\nabla{T}, 
\label{Drude-Fourier}
\end{equation}

\noindent
where $v^2$ is the electron mean square velocity and $c_V$ is the electron specific heat.  By comparing Equation (\ref{Drude-Fourier}) with Equation (\ref{Fourier}), the thermal conductivity $\kappa$ can be identified as

\begin{equation}
\kappa = \frac{1}{3}v^2 \tau c_V. 
\label{Drude-Thermal}
\end{equation}

\noindent
Because Drude envisioned electrons as particles of classical gas, their mean kinetic energy is related to the temperature via the equipartition theorem,

\begin{equation}
\frac{1}{2}m v^2 = \frac{3}{2} k_B T,
\label{Drude-Kinetic}
\end{equation}

\noindent
from which the expression for the average velocity of electrons can be readily derived,

\begin{equation}
v = \sqrt{\frac{3k_B T}{m}},
\label{Drude-Velocity}
\end{equation}

\noindent
where $k_B = 8.62 \times 10^{-5}$ eV K$^{-1}$ is the Boltzmann constant. Similarly, the classical heat capacity of an ideal gas with particle density $n$ reads

\begin{equation}
c_V = \frac{3}{2} n k_B,
\label{Drude-Capacity}
\end{equation}

\noindent
since each degree of freedom contributes $\frac{1}{2} k_B$. Using the expressions of the thermal and electric conductivities given in Equation (\ref{Drude-Thermal}) and Equation (\ref{sigma}), respectively, we can evaluate their ratio for a given temperature, that is, the Lorenz number given in Equation (\ref{Lorenz}),
\begin{equation}
L = \frac{\kappa}{\sigma} \cdot \frac{1}{T}  = \frac{1}{3}v^2 \tau c_V \cdot  \frac{m}{ne^2\tau} \cdot \frac{1}{T}.
%L = \frac{\kappa}{\sigma T} = \frac{1}{3} v^2 c_V \cdot \frac{m}{ne^2 T}.
\label{Drude-Lorenz}
\end{equation}
By replacing $c_V$ with the expression in Equation (\ref{Drude-Capacity}) and $v$  with the expression in Equation (\ref{Drude-Velocity}), we obtain the Lorenz number as predicted by Drude's theory,
\begin{equation}
L =  \frac{3}{2} \ \left(\frac{k_B}{e}\right)^2 = 1.10  \times 10^{-8} \ \textnormal{W} \ \Omega / \textnormal{K}^2.
\label{Drude-Lorenz-2}
\end{equation}

\noindent
From this result, we conclude that Drude's theory successfully explains the empirical Wiedemann-Franz law by correctly predicting that the Lorenz number is (i) metal-independent, (ii) temperature-independent, and (iii) on the order of $10^{-8}$ W ${\Omega}$ / K$^2$, differing from typical experimental values by less than 0.00000002 W ${\Omega}$ / K$^2$.\footnote{Curiously, in the original derivation Drude introduced an error of a factor of two, leading to a Lorenz number twice as large ($2.20  \times 10^{-8} \ \textnormal{W} \ \Omega / \textnormal{K}^2$) and in closer agreement with empirical data, as discussed by \cite{Singh2024}. Of course, this error has no scientific or philosophical significance.} The comparison between the prediction of Drude's theory and experimentally observed values of the Lorenz number is presented in Figure \ref{Figure}(a). As noted by  \cite{Simon2013}, ``this result was viewed as a huge success,'' in that ``before Drude no one had any idea why this ratio should be a constant at all!'' (p.\ 23).

\section{Understanding the Success of Drude's theory} 
\label{Sec-3} 
As we have seen, Drude's theory offers a microscopic account of the Wiedemann-Franz law through a quantitative prediction of the Lorenz number. What is the origin of this empirical success? Does it warrant belief, as the realist suggests, that electrons in metals behave as described by Drude? And if not, what sort of antirealist explanation might we appeal to in order to make sense of this success?

\paragraph{\emph{Drude and truth.}} Realists often consider novel predictions as the hallmark of the empirical success of scientific theories. Various accounts of what constitutes a novel prediction have been proposed. I will adopt the `heuristic view' developed by \cite{Worrall1985, Worrall1989} and endorsed by realists such as \cite{Musgrave1988}. According to this view, novelty is not strictly determined by the temporal order in which a theory and evidence emerge. Instead, a prediction is considered novel if the predicted fact was not directly used in constructing the theory. As discussed earlier, Drude developed his theory without relying on any assumptions regarding electrical or thermal conductivity. His prediction of the Lorenz number, in quantitative agreement with the empirical observation, therefore qualifies as novel.

Drude's theory thus stands as a mature scientific theory that is predictively and explanatorily successful. This should license the realist to believe that the theory portrays an approximately true description of electrons in metals: ``The basic idea of the realist,'' according to \cite{Vickers2019},  ``is this: where science is sufficiently successful---makes accurate predictions and/or exhibits significant explanatory power---the relevant theoretical hypotheses are (probably and/or approximately) true'' (p.\ 571). Returning to the words of \cite{Putnam1978}, whenever theories consisting of ``electron calculi,'' of the sort developed by Drude, ``correctly predict observable phenomena,'' such as the Lorenz number, ``then a natural explanation of the success of these theories is that they are partially true accounts of how [electrons] behave.'' (pp.\ 18-19).  Indeed, ``in Drude’s time,'' as observed by \cite{Hofmann2015}, ``one of the most convincing pieces of evidence for his theory appeared to be that it yielded a quantitatively correct description of the Wiedemann-Franz law.'' (p.\ 85). Similarly, \cite{Alloul2011} emphasizes that this empirical success ``was long regarded as a proof that Drude’s hypothesis as
formulated in the context of classical mechanics was actually correct'' (p.\ 107).

Is Drude's theory true, then? According to Drude's theory, electrons in metals behave like an ideal gas of classical particles: they move freely, occasionally colliding with immobile ionic cores, while neither interacting with one another nor with the ionic cores beyond these collisions. This picture, however, starkly contrasts with our present understanding. In the face of our current knowledge, Drude's description of electrons in metals is unequivocally false. Electrons are not classical particles, as their De Broglie wavelength ($\sim$$0.5$ nm) is comparable to interatomic distances in metals ($\sim$$0.3$ nm), thus requiring an account of their quantum nature. Nor can they be viewed as an ideal gas, since their density ($\sim$$10^{22}$ cm$^{-3}$) exceeds that of a classical gas at standard temperature and pressure ($\sim$$10^{19}$ cm$^{-3}$)  by three orders of magnitude. Furthermore, electrons do not undergo collisions with ions, nor do they scatter with a constant mean free time. Instead, they scatter off lattice vibrations (phonons) and impurities---processes entirely absent from Drude’s framework. Finally, electron-electron and electron-ion electromagnetic interactions can be sizable (several eV), with the attractive potential created by the periodically arranged ionic cores playing a crucial role in determining the behavior of electrons, as described by Bloch's theorem. Despite its empirical success, Drude's theory is a far cry from an approximately true account of electrons in metals.

The realist may object that, even though Drude's theory says only false things about electrons, it is nevertheless successful in referring to such real entities. Surely electrons exist in metals---that much is uncontroversial according to present science---aligning with \cite{Putnam1975}'s claim that   ``terms in a mature science typically refer'' (p.\ 73). Is referential success sufficient to regard a theory as approximately true? The relationship between truth and referential success has been examined by, e.g.,  \cite{Musgrave1988}, who correctly observed that  ``reference, while a necessary condition for truth, is not a sufficient condition'' given that  ``a theory may be referential yet false.'' (p.\ 236). To illustrate this, consider the following claim about the real person named Michele Pizzochero: `Michele Pizzochero is ten meters tall.' The statement successfully refers---there is, in fact, a person named Michele Pizzochero---but it fails to describe him, and is thus false.\footnote{This example is loosely inspired by  \cite{Musgrave1988}'s passage: ``Richard Nixon is tall, blonde, honest and never swears.'' According to \cite{Musgrave1988}, this statement ``refers to Richard Nixon all right, but it says a lot of false things about him.'' (p.\ 236).} Analogously, the claim `Electrons are classical particles' that underlies Drude's theory is false, albeit referentially successful. Drude's theory is, in \cite{Musgrave1988}'s account, one of ``the  \emph{false} theories of our ancestors'' even though ``referred to electrons just as our own theories do'' (p.\ 236, emphasis added). However, \cite{Musgrave1988} continues, ``in any realist explanation of science's success it is truth or near-truth which is going to be important, rather than mere successful reference'' (p.\ 236).

\paragraph{\emph{Drude and fluke.}}

To gain insight into the origin of the success of Drude's theory in quantitatively accounting for the Wiedemann-Franz law, it is instructive to consider Sommerfeld's theory,\footnote{Not to be confused with Sommerfeld's theory of fine structure discussed in Section \ref{Sec-5}.} which we will regard as the most successful---and, from a realist perspective, the closest to truth---account of electrons in metals.\footnote{By denoting Sommerfeld's theory as ``true'' or ``closest to truth,'' I do not mean to endorse it as literally true in any naïve realist sense. The label is meant in a sense \emph{relative} to the prior Drude's theory: Sommerfeld’s framework is empirically more successful, and it is grounded in the quantum-mechanical description of electrons that remains central to contemporary condensed matter physics. For this reason, the realist may reasonably treat it as a case where empirical success and approximate truth coincide. My broader point, however, is that such realist classification is not always reliable. As Drude's theory illustrates, empirical success can arise for reasons other than approximate truth.} By comparing the Lorenz number predicted by Sommerfeld's theory with that of Drude's theory, I will argue that the success of the latter is the result of a fortunate coincidence.

Nearly thirty years after the work of Drude, \cite{Sommerfeld1928} was the first to apply the ideas of quantum mechanics to metallic conduction.\footnote{For a detailed presentation of Sommerfeld's theory, see, e.g., Chap.\ 2 of \cite{Ashcroft1976}, Chap.\ 6 of \cite{Kittel1986}, Chap.\ 4 of \cite{Alloul2011}, Chap.\ 4 of \cite{Simon2013}, Chap.\ 1 of \cite{Singleton2001}, and Chap.\ 6  of \cite{Hofmann2015}.} Specifically, Sommerfeld's theory reflects the quantum nature of electrons in metals by describing them as a gas of free fermions, thus obeying the Fermi-Dirac statistics which had been developed just a couple of years earlier,
\begin{equation}
f(\epsilon, T) = \frac{1}{e^{( \epsilon-\mu)/(k_BT)}+1},
\end{equation}
\noindent
where $f$ is  the distribution function which gives the probability of occupation of an electronic state of energy $\epsilon$ at temperature $T$ and $\mu$ is the chemical potential. To obtain the ground state of the $N$ free electrons confined in, e.g., a cubic metal of length $L$ and volume $V = L^3$, one needs to determine the set of energy levels $\epsilon$  and fill them in order of increasing energy and in accordance with the Pauli exclusion principle. The starting point is the time-independent Schr\"oedinger equation for the free electron (i.e., in which only the kinetic energy operator appears in the Hamiltonian) subject to Born-von Karman boundary conditions on each of the three spatial coordinates of the wave-function $\psi$, i.e., $\psi(x,y,z) = \psi(x+L, y, z)$ for the $x$ direction, and analogously for $y$ and $z$. 

The resulting electron energy is
\begin{equation}
\epsilon(\vec{k}) =  \frac{\hbar  \vec{k}^2}{2 m}
\label{energy}
\end{equation}
where $\hbar = 6.58 \times 10^{-16}$ eV s is the reduced Planck constant and $\vec{k}$ is the quantized electron wave-vector, the components of which take the form
\begin{equation}
k_x = \frac{2\pi}{L}n_x, \qquad k_y = \frac{2\pi}{L}n_y, \qquad k_z = \frac{2\pi}{L}n_z,
\end{equation}
with $n_x$, $n_y$, and $n_z$  being integer numbers. In the three-dimensional $\vec{k}$-space of axes $k_x$, $k_y$, and $k_z$, the allowed wave-vectors are equally spaced and each corresponding to an energy level, according to Equation (\ref{energy}). The $N$-electron ground state can be constructed by allocating electrons into the allowed energy levels specified by the wave-vector $\vec{k}$. To that end, at most two electrons with opposite spins should be placed in each level. First, the energy level at $\vec{k} = 0$, which has vanishing energy, is populated. Next, the remaining electrons are added to the otherwise empty levels in ascending energy. Once all the $N$ electrons are accommodated, the region encompassing the occupied levels in $\vec{k}$-space forms a sphere, the so-called Fermi sphere. The radius of the Fermi sphere is the Fermi wave-vector, $k_F$, which depends on the number of electrons $N$ and the crystal volume $V$ through their ratio, the electron density $n = N/V$, as
\begin{equation}
k_F = \sqrt[3]{3 \pi^2 n}.
\end{equation}
which is typically on the order of $0.1$ nm$^{-1}$. The energy of the electrons with wave-vector $k_F$ is the Fermi energy, 
\begin{equation}
\epsilon_F =  \frac{ \hbar {k^2_F} }{2 m},
\label{Fermienergy}
\end{equation}
which is the energy of the most energetic electrons in the metal, on the order of 1-10 eV. Because electrons in Sommerfeld's theory are non-interacting, this energy is solely kinetic energy,
\begin{equation}
\epsilon_F =  \frac{1}{2}mv_F^2,
\end{equation}
from which the Fermi velocity $v_F$ can be readily derived,
\begin{equation}
v_F = \sqrt{\frac{2 \epsilon_F}{m}}.
 \label{quantum-vel}
 \end{equation}
 By considering the Fermi-Dirac statistics in the case of the free electron gas, the electron specific heat is
\begin{equation}
c_V =\frac{\pi^2}{2} \left(\frac{k_BT}{\epsilon_F}\right) n k_B. 
\label{quantum-hc}
 \end{equation}
 Using Equation (\ref{Drude-Lorenz}), we can determine the Lorenz number using the quantum electron velocity and heat capacity given in Equation (\ref{quantum-vel}) and Equation  (\ref{quantum-hc}), respectively, derived from Sommerfeld's theory, instead of the classical ones from Drude's theory,
\begin{equation}
 L =  \frac{ \pi^2}{3} \ \left(\frac{k_B}{e}\right)^2 = 2.44  \times 10^{-8} \ \textnormal{W} \ \Omega / \textnormal{K}^2.
 \label{Sommerfeld-Lorenz}
 \end{equation}
 This result is remarkably close to that obtained using Drude's theory in Equation (\ref{Drude-Lorenz-2}) and, again, in  agreement with empirical observations, being (i) metal-independent, (ii) temperature-independent, and (iii) in excellent accord with experimental values, as shown in Figure \ref{Figure}(a).
 
With Sommerfeld's theory at hand, we are now in a position to explain the cause of the empirical success of Drude's theory. As is evident from Equation (\ref{Drude-Lorenz}), the Lorenz number depends on the product of the electron velocity squared and heat capacity,
 \begin{equation}
 L  \propto v^2 c_V,
 \label{propto}  
 \end{equation}
 with all other terms being constant or canceling out. It is revealing to compare the electron velocity and heat capacity according to the `true` Sommerfeld's theory and the `false' Drude theories for typical metals at room temperature. Sommerfeld's theory yields $v_F \approx 10^6$ m/s, while Drude's theory yields $v \approx 10^5$ m/s. Hence, Drude's theory underestimates the electron velocity by one order of magnitude. With respect to the electron heat capacity,  Sommerfeld's theory yields $c_V \approx10^{-7}$ eV/K per electron, while Drude's theory yields $c_V \approx 10^{-5}$ eV/K per electron. Hence, Drude's theory overestimates the electron heat capacity by two orders of magnitude. More rigorously, by comparing Equation (\ref{quantum-vel}) and Equation (\ref{Drude-Velocity}), one notices that the square of the Fermi velocity in Sommerfeld's theory is larger than the thermal mean square speed in Drude's theory by a factor $\sim$$\frac{\epsilon_F}{k_BT}  \approx 100$ at room temperature. Similarly, by comparing Equation (\ref{quantum-hc}) and Equation (\ref{Drude-Capacity}), one notices that the heat capacity in Sommerfeld's theory is smaller than Drude's theory by a factor $\sim$$\frac{k_BT}{\epsilon_F}  \approx 0.01$ at room temperature. In addition to these quantitative differences in electron velocity and heat capacity, there is also an important qualitative difference. In Drude's theory, the electron heat capacity is temperature-independent, while electron velocity is temperature-dependent; cf.\ Equation (\ref{Drude-Capacity}) and Equation (\ref{Drude-Velocity}).  Conversely, in Sommerfeld's theory, the electron heat capacity is temperature-dependent, while the electron velocity (i.e., the Fermi velocity) is temperature-independent; cf.\ Equation (\ref{quantum-hc}) and Equation (\ref{quantum-vel}).

This is where the fluke comes into play. Because the Lorenz number arises from the  \emph{product} of the square of the electron velocity and the heat capacity, as indicated in Equation (\ref{propto}), Drude's theory ends up benefiting from  a compensation of errors: it underestimated the square of the electron velocity by two orders of magnitude while overestimating the heat capacity by the same amount---and the two roughly cancel out in their product. Furthermore, although Drude's theory assumes that heat capacity is temperature-independent while electron velocity depends on temperature---the opposite of what Sommerfeld's theory correctly predicts---this wrong temperature dependence ultimately cancels out too. As a result, Drude's theory successfully predicts the Lorenz number not by virtue of an approximately true description of the underlying physical reality, but from a cancellation of errors that is purely accidental. This is widely acknowledged in the condensed matter physics literature. For example, \cite{Simon2013} emphasizes that Drude's ``calculation is completely incorrect (despite its successful result), but the reason why ``this calculation gives such a good result'' is due to ``two mistakes that roughly cancel each other. We have used a specific heat that is way too large, but we also used a velocity that is way too small.'' (p.\ 23). \cite{Ashcroft1976} observe that the ``Drude's {fortuitously} good value'' is ``thanks to the two compensating corrections of order $\frac{\epsilon_F}{k_BT}$'' (p.\ 52). \cite{Singh2016} is firm about why the ``Drude's theory was able to predict the correct value of the Lorenz number,'' namely, ``\emph{just a coincidence}'' (p.\ 11, emphasis added). Similarly, \cite{Alloul2011} characterizes the success of Drude's theory as an ``\emph{amazing coincidence}'' (p.\ 111, emphasis added).   \emph{Pace }\cite{Smart1963}.

\section{Three Drude-imentary Lessons} 
\label{Sec-4} 
Drude's theory succeeds in accounting for the Wiedemann-Franz law not by telling the true story of electrons in metals, but because large errors in electron velocity and heat capacity, along with their temperature dependence, happen to cancel each other out. From the realist perspective, this lucky accident leads to the right result for the wrong reasons. Though one should be cautious about drawing broad conclusions from a single case, I suggest we can learn three important lessons for the scientific realism debate.

\paragraph{\emph{Lesson \#1: Coincidence is a legitimate explanation for scientific success.}}  Drude's theory scores empirical success in explaining a phenomenon at the observational level through a cancellation of errors at the unobservable level. Thus, we have a mechanism through which coincidences can be responsible for the empirical success of a scientific theory: offsetting falsehoods. This casts doubt on the generality of \cite{Smart1963}'s view, according to which the interpretation of   ``a theory in a realist way'' eliminates ``the need for such a cosmic coincidence''  (p.\ 39). Drude's theory is plainly false by present lights---electrons in metals just do not behave as Drude envisioned---but it is precisely a coincidence that explains its correct prediction of the Lorenz number. Of course, this episode does not imply that coincidence is a universal explanation for scientific success, nor does it rule out that approximate truth can be the source of the success of certain theories. Nevertheless, it illustrates that coincidences, such as those arising in the form of error compensations, offer a   plausible yet overlooked explanation for the success of a scientific theory.  We possess no definitive proof, beyond philosophical speculation, that empirical success is the signature of truth, given that the unobservable reality lies beyond our epistemic reach.\footnote{\cite{Chang:2022} challenges `correspondence realism' on the basis of what he calls the fallacy of pre-configuration, i.e., the assumption that ``reality has well-defined parts and properties that exist independently of all conceptualization'' (p.\ 74).} In contrast, the  success of Drude's theory in accounting for  the Lorenz number stands as concrete evidence that the success of what is now regarded as a false theory can occur through a lucky accident.

\paragraph{\emph{Lesson \#2: Empirical success may not scale with truth.}} The realist claims that the empirical success of a scientific theory is best explained by its approximate truth. On this basis, one may expect that an increase in the approximate truth invariably leads to an increase in empirical success. 
Drude's theory predicts a Lorenz number that deviates from the value empirically observed by only a factor of 2, despite having both a false electron velocity and heat capacity, each being incorrect by a factor of 100 at room temperature. On the other hand, Sommerfeld's theory---which we accept as the true account of the subject matter---yields a prediction of the Lorenz number that matches the empirically observed value. But would an increase in the truth content of Drude's theory translate to an increased empirical success, i.e., a prediction of a Lorenz number that is closer to the observed value? To quantitatively answer this question, let us consider a thought experiment involving a fictitious version of Drude's theory. In this fictitious Drude's theory, we assume the `true' electron velocity (i.e., the Fermi velocity from Sommerfeld's theory)  while retaining the heat capacity of the original Drude's theory. This version of the theory enjoys a higher truth content than original Drude's theory. Yet,  the prediction of the Lorenz number is markedly worse than that of the original Drude's theory, overestimating its value by two orders of magnitude. On the other hand, consider a scenario where the fictitious Drude's theory adopts the `true' heat capacity (i.e., the one predicted by Sommerfeld's theory), but retains the electron velocity of the original Drude's theory. In this case, the predicted Lorenz number would be underestimated by two orders of magnitude. Both fictitious versions of Drude's theory achieve worse empirical success than the original one, despite featuring a higher degree of truth. This is illustrated in Figure \ref{Figure}(c), which compares the predictions of the Lorenz number obtained with the original Drude's theory, the fictitious variants, and Sommerfeld's theory, alongside the empirically observed value for the representative case of silver at room temperature. Furthermore, both fictitious Drude theories would lead to a ratio of the thermal conductivity to the electrical conductivity that is not linear in the temperature, unlike the  Wiedemann-Franz law given in Equation (\ref{Lorenz}). This thought experiment quantifies an important point: an increase in truth content in a false theory does not necessarily entail an increased empirical success.

%@article{Weston:1992,
 %author = {Thomas Weston},
 %journal = {Philosophy of Science},
 %number = {1},
 %pages = {53--74},
 %title = {Approximate Truth and Scientific Realism},
 %volume = {59},
 %year = {1992}
%}

\paragraph{\emph{Lesson \#3: Antirealist explanations for scientific success can work in synergy. }} The realist provides a single explanation for the success of a scientific theory: approximate truth. In contrast, antirealists can appeal to multiple  explanations that may complement one another to shed light on what makes a scientific theory successful. For instance, consider \cite{Stanford2000}'s antirealist account based on predictive similarity, according to which ``the success of a given false theory in a particular domain is explained by the fact that its predictions are (sufficiently) close to those made by the true theoretical account of the relevant domain'' (p.\ 275). Stanford’s view provides a plausible explanation for the success of Drude's theory.\footnote{I do not aim to endorse or criticize \cite{Stanford2000}'s account. My purpose here is to demonstrate that it can be straightforwardly applied to explain the success of the false Drude's theory in light of the true Sommerfeld's theory.} The predictions of Drude's theory are indeed very close to those made by the true Sommerfeld's theory: both theories predict a Lorenz number $L \propto \left(\frac{k_B}{e}\right)^2$, with the two differing by less than 0.00000002 W ${\Omega}$ / K$^2$, as can be seen by comparing
Equation (\ref{Sommerfeld-Lorenz}) and  Equation (\ref{Drude-Lorenz}). The explanation based on predictive similarity does not conflict with an explanation based on coincidence. In fact, the two explanations can work together, with the latter deepening the former.\footnote{The coincidental relationship between Drude and Sommerfeld theories may perhaps address some of \cite{Psillos2001}'s criticisms of \cite{Stanford2000}'s proposal, demonstrating that predictive similarity can support a deeper explanation without requiring a realist commitment to theories.}  That is, the predictive similarity of Drude's theory to Sommerfeld's theory occurs via a  coincidence. This suggests that the antirealist may draw on multiple explanations to offer a robust account of how a scientific theory achieves empirical success---an argumentative advantage not available to the realist.

\section{Drude's theory: Like or Unlike the Sommerfeld Puzzle?} 
\label{Sec-5} 
It is instructive to set the success of Drude's theory alongside another episode in twentieth-century physics where coincidence is thought to have played a decisive role: Sommerfeld’s prediction of the fine-structure formula. Physicists widely regard Sommerfeld’s success\footnote{Not to be confused with Sommerfeld's theory of electrons in metals discussed in Section \ref{Sec-3}.} as a fortunate accident, akin to Drude’s prediction of the Lorenz number. However, a recent philosophical analysis by \cite{Vickers2020} has offered an alternative reading of Sommerfeld's formula that is compatible with a selective form of scientific realism. I will suggest, however, that extending this line of reasoning to Drude's theory may present far greater difficulties.

\paragraph{\emph{The Sommerfeld puzzle.}} Every chemical element in the periodic table emits and absorbs electromagnetic radiation only at specific frequencies, collectively known as the spectrum. The first theoretical framework of the spectrum of hydrogenic atoms (i.e., one-electron atoms such as hydrogen) was developed within the old quantum theory. In 1913, \cite{Bohr1913a, Bohr2013b}  derived the frequency of the spectral lines by assuming that an electron in an atom can only move in discrete, circular orbits around the nucleus, with its energy $E_n$ being quantized by the principal quantum number $n = 1, 2, 3, ...$ as $E_n = \frac{hR}{n^2}$, where $h$ is the Planck constant and $R$ the Rydberg constant. The frequency $\nu$ (and thus the energy, according to Einstein's formula $E = h\nu$) of the radiation emitted is due to quantum jumps, i.e., abrupt electron transitions from one allowed orbit (e.g., $n_2$) to another that lies lower in energy (e.g., $n_1$), as
 \begin{equation}
  \nu  = R    \left (\frac{1}{n_2^2} - \frac{1}{n_1^2}   \right).
 \end{equation}
Bohr achieved notable empirical success in describing the spectrum of hydrogen. Yet it soon became clear that such success was incomplete. In 1887, Michelson and Morley revealed that certain spectral lines are split into multiple sublevels---an effect known as the ``fine structure'' of the spectrum.

Three years later,  \cite{Sommerfeld1916} built upon Bohr's theory to derive a formula which accurately predicted the fine structure of hydrogen. Sommerfeld's reasoning proceeded in two main steps. First, he posited that for every allowed energy within Bohr's theory, there should be an infinite number of possible orbits, reflecting the infinite range of elliptical eccentricities that could characterize the trajectory of electrons. However, electrons are restricted to certain elliptic orbits, an effect described by a second quantum number $k$, in addition to the principal quantum number $n$ previously introduced by Bohr. Second, Sommerfeld applied the laws of special relativity to the electron trajectories, according to which the electron mass changes with its velocity. The presence of the two quantum numbers $n$ and $k$ lifts the energy degeneracy of the spectral lines, leading to the emergence of the fine structure. Overall, by relying on relativistic classical mechanics, Sommerfeld derived the following expression for the allowed energies (and hence, the frequencies) of the hydrogenic atom,

 \begin{equation}
  E(n, k)  = - \frac{RhcZ^2}{n^2}  \left[1+  \frac{\alpha^2 Z^2}{n^2}  \left( \frac{n}{k}- \frac{3}{4} \right) \right] + ...
 \end{equation}
 
 \noindent
where $c$ is the speed of light, $Z$ is the atomic number, and $\alpha$ is the fine structure constant. The Sommerfeld formula, as \cite{Vickers2020} notes, ``encodes countless novel predictions of spectral lines with extreme quantitative accuracy'' (p.\ 989) and thus prompts the realist commitment to, e.g., the existence of quantized elliptical orbits. Here lies the puzzle. On the one hand, Sommerfeld's formula rests on a radically false picture of reality, if judged from the standpoint of contemporary physics: elliptic orbital trajectories have no place in the modern quantum theory, with the fine structure originating from the electron spin and its interaction with orbital motion (that is, spin-orbit coupling), a feature that is entirely absent from Sommerfeld’s theory. On the other hand, Sommerfeld nevertheless derived the same exact formula that would later arise from Dirac’s relativistic quantum mechanics of 1928, a result that remains in use to this day. How can we explain this ``remarkable twist of historical fate,'' as  \cite{Barley2025} (p.\ 15) calls it, by which Sommerfeld managed to derive his correct formula in 1916, despite modern quantum theory and electron spin still lying a decade in the future?

\paragraph{\emph{Coincidental explanations of Sommerfeld's formula.}} Many prominent figures in theoretical physics are adamant that Sommerfeld's prediction is essentially a fluke. To cite just a few examples, it was characterized as ``a miracle'' by \cite{Heisenberg1968} (p.\ 534), deemed ``only fortuitous'' by Schrödinger in his 1956 letter to \cite{Yourgrau1960} (p.\ 114), labeled an ``accident'' by \cite{Weinberg1995} (p.\ 6), described as ``a coincidence'' by \cite{Eisberg1985}, and perhaps even ``the most remarkable numerical coincidence in the history of physics'' by \cite{Kronig1960} (p.\ 9). Such characterizations fly in the face of the arguments of \cite{Smart1963, Smart1968} and \cite{Putnam1978} for scientific realism.

Yet, as argued in Section \ref{Sec-1}, coincidences call for an explanation. Several accounts have been proposed to understand the coincidental success of Sommerfeld's formula. For instance,  \cite{Yourgrau1960}  submit that ``Sommerfeld's explanation was successful because the neglect of wave mechanics and the neglect of spin by chance cancel each other in the case of the hydrogen atom'' (p.\ 115). Analogously, \cite{Eisberg1985} maintain that the ``coincidence occurs because the errors made by the Sommerfeld model, in ignoring the spin-orbit interaction and in using classical mechanics to evaluate the
average energy shift due to the relativistic dependence of mass on velocity, happen to cancel for the case of the hydrogen atom'' (p.\ 286). This qualitative idea has been translated onto a rigorous footing by \cite{Keppeler2003}, who, through elaborate analytical calculations, demonstrate ``why Sommerfeld  was able to derive the correct energy eigenvalues before the development of quantum mechanics and before the discovery of spin [...]: the contribution deriving from the Maslov index and the influence of spin accidentally cancel each other'' (p.\ 86).

The picture that emerges from this analysis of Sommerfeld's prediction of the fine structure is, \emph{mutatis mutandis}, analogous to Drude's prediction of the Lorenz number: two mature scientific theories that achieve predictive success in their respective domains through coincidences stemming from fortuitous cancellations of errors, while portraying images of reality that are false by comparison with our current science. If the arguments concerning the accidental character of Sommerfeld’s prediction of the fine structure are accepted, then the three lessons outlined in Section \ref{Sec-4} receive further support. First, coincidence---in the form, for example, of error compensation---constitutes a legitimate yet truth-independent explanation of scientific success. Sommerfeld's account of reality is partially non-referential, insofar as it relies on elliptic orbits while neglecting key features of modern quantum theory, such as electron spin. Second, an increase in the truth-content of Sommerfeld’s theory would not necessarily have yielded greater predictive accuracy. Had Sommerfeld incorporated electron spin or the wave-like character of electrons—as encoded in modern quantum mechanics—the compensatory error mechanism would have vanished. Third, the predictive success of Sommerfeld’s theory can also be understood from an antirealist perspective, for instance, by appealing to \cite{Stanford2000}'s predictive similarity to the fine-structure results of Dirac's relativistic quantum mechanics. \cite{Vickers2012} argues for an antirealist reading of Sommerfeld's prediction of the fine structure, insisting that it ``is almost certainly impossible'' (p.\ 1) for the realist to explain its success in terms of approximate truth.

\paragraph{\emph{Can structural realism come to the rescue?}} In a subsequent essay, however, \cite{Vickers2020} develops an interpretation of the transition from Sommerfeld’s to Dirac’s theory compatible with a more sophisticated variant of scientific realism, moving beyond the previously discussed ``two-errors-canceling-out'' explanation. To be sure, \cite{Vickers2020} concedes that, if judged through the lens of standard scientific realism, Sommerfeld’s theory is problematic: ``elliptical orbits are not involved in the Dirac quantum mechanics, and electron spin is not even approximately involved in Sommerfeld’s theory'' (p.\ 995). In addition, there remain profound differences between the framework of classical mechanics, which underpins Sommerfeld’s approach, and that of quantum mechanics, which anchors Dirac’s.

Yet, \cite{Vickers2020} argues that the apparent gulf between the two theories can be bridged by appealing to a form of selective scientific realism, which restricts realist commitment to those theoretical components genuinely responsible for empirical success. In particular, \cite{Vickers2020} maintains that ``Sommerfeld's success is down to the fact that his theory includes sufficient structural truth'' (p.\ 1006). Building on the detailed mathematical analysis of \cite{Biedenharn1983}, he further suggests that ``It is clear that Sommerfeld didn't need to get the physics exactly right, so long as the structure of his theory took on a certain form. But more importantly, Sommerfeld didn't even need to get the structure right, in two different ways. First, we need to make a distinction between `working' and `idle' structure (cf.\ \cite{Votsis2010}). Sommerfeld used the BWS quantum conditions, but these are more specific than they need to be […]. Sommerfeld's derivation is just as successful with the more abstract conditions […] (that is, staying neutral on BWS and semi-classical conditions). Second, Sommerfeld's theory was missing some structure that is a crucial part of Dirac QM. But it turns out that the missing structure is idle vis-à-vis the fine structure formula, so it didn't matter that Sommerfeld missed it out'' (p.\ 1006). Thus, Sommerfeld’s achievement lay not in correctly representing the underlying physical reality, but in encoding just enough of the relevant structural features to secure predictive accuracy in the hydrogen case. \cite{Vickers2020}’s philosophical analysis lends support to structural realism---the view, revived by \cite{Worrall1989a} from the earlier insights of \cite{Poincare1902, Poincare1905}, according to which science delivers knowledge of the relational structure of reality rather than of its intrinsic content, with the former being preserved across theory change while the latter being subject to discontinuity.

It is natural to ask whether a structural realist approach could be applied to Drude's theory by identifying sufficient continuity with Sommerfeld's theory of electrons in metals in the prediction of the Lorenz number. Such a demonstration would be essential for endorsing a structural realist reading of this episode. In the absence of a no-go theorem, it would be premature to declare this impossible. However, the task seems more daunting than in the case of Sommerfeld’s fine-structure formula. First, in contrast with the fine-structure case where \cite{Vickers2020} could rely on the calculations of \cite{Biedenharn1983}, the only explanation currently available for Drude's theory in the physics literature appeals to a coincidental success via error cancellation. Second, unlike the fine-structure case, the relevant equations in Drude's theory are not preserved intact across the transition to Sommerfeld's theory. This indicates that any structural retention is partial and far less evident than in the fine-structure case. Third, the structural differences within each theory are profound, particularly with respect to temperature dependence. According to Drude, the electron heat capacity is temperature-independent while the electron velocity varies with temperature, whereas Sommerfeld predicts the opposite. The predictive success of Drude thus relies entirely on this compensatory interplay, suggesting that any attempt to recover a structural realist continuity may face  obstacles. The structural realist has the floor.

\section{Two Wrongs Make One Right} 
\label{Sec-6} 
\paragraph{\emph{The challenge to scientific realism.}} Having established that a coincidence---in the form of two errors canceling out---provides the best current explanation for the success of Drude's theory in predicting the Lorenz number, it is important to assess the extent to which this episode challenges scientific realism. At first glance, this coincidental success appears to contradict \cite{Smart1963, Smart1968}'s argument. Yet the realist may respond that scientific realism is best understood as a population-level view concerning scientific theories in general: the success-to-truth inference underlying \cite{Smart1963, Smart1968}'s and \cite{Putnam1975, Putnam1978}'s arguments is to be construed as a claim about the overarching pattern of scientific success across history, rather than about the fate of any single theory in isolation. Under this framework, Drude's theory constitutes only a single counterexample that does not undermine the overall pattern. It is an anomaly within a largely stable trend---akin to a student who attains a high score by cheating, which does not threaten the general conclusion that students’ high scores are best explained by their understanding.\footnote{I am grateful to a reviewer for raising this objection.}

I argue that dismissing examples and counterexamples is a weak argumentative strategy for the realist. The debate over scientific realism is more fruitful when it engages directly with specific theories and historical episodes, working at the `local' level. In defending a realist interpretation of Sommerfeld’s fine-structure episode---though the same point applies equally to Drude's theory---\cite{Vickers2020} aptly observes that

\begin{quote}
Many figures in the contemporary realism debate do not like to talk in terms of counterexamples to realism. Instead realism is said to involve a defeasible commitment: scientific success is (highly) indicative of truth, but does not guarantee it. But this overlooks the fact that the measure of scientific success, including predictive success, is a matter of degree (cf.\ \cite{Fahrbach2011}). The more impressive the success, the closer the realist comes to inferring that success must be born of truth. Thus one can see why some have been tempted to refer to the Sommerfeld case as a `counterexample', at least loosely speaking: the quantitative accuracy of Sommerfeld’s fine structure formula is extremely impressive, and thus highly motivating for the realist. Thus it seems especially hard in this case for the realist to shrug her shoulders and say `Well, this is just one case, and my inference is defeasible' (p.\ 988).
\end{quote}

Furthermore, one of the most influential realist responses to \cite{Laudan1981}’s pessimistic meta-induction---which draws on the historical record of specific, once-successful but now-abandoned scientific theories---has been the development of selective forms of realism, such as \cite{Psillos:1999}'s deployment realism or \cite{Worrall1989}'s structural realism, to mention only two representative approaches. Although these positions differ in important respects, they share the conviction that defending a realist interpretation of theory change requires a case-by-case analysis.\footnote{``In the meantime, perhaps influenced by Laudan’s focus on individual theories,'' \cite{Vickers2020} argues, ``the discussion gradually shifted away from a ‘global’ focus on the success of science (e.g. Putnam, Boyd) and towards a ‘local’ focus on the success of individual theories.'' (p.\ 574)} Each theory must be examined in detail to distinguish between constituents that are merely idle and those genuinely responsible for its predictive success---ideally, the ones preserved through theory change. This approach underlies \cite{Psillos:1999}’s interpretation of the transition from the caloric theory of heat to thermodynamics and \cite{Worrall1989}’s reading of the shift from the solid ether theory to Maxwell’s theory of the electromagnetic field---both examples drawn from \cite{Laudan1981}’s list. Likewise, \cite{Vickers2020} employs a similar structural realist strategy in his proposed solution to the Sommerfeld puzzle.

Closely related, if one aims to move beyond ``armchair'' metaphysics toward a ``naturalized'' metaphysics, where metaphysical commitments are informed and constrained by our best science, then a population-level conception of realism risks becoming vacuous. The broad assertion that ``most scientific theories are successful because they are approximately true'' is too coarse-grained to be illuminating. What is required instead is a narrowly focused analysis, one where our metaphysical picture of reality is guided by the most successful and stable constituents of scientific theories within their respective domains.

\paragraph{\emph{Concluding remarks.}} In summary, I have examined the empirical success of Drude's theory in accounting for the Wiedemann-Franz law through the lens of the scientific realism debate. I have argued that Drude's theory does not predict a Lorenz number---the quantity that governs the Wiedemann-Franz law---by virtue of its approximately true description of the unobservable reality, that is, the behavior of electrons in metals. Instead, the empirical success stems from a coincidence:  a fortunate compensation of errors involving an overestimation of electron velocity and  an underestimation of electron heat capacity, along with their temperature dependence.

From this episode, I have suggested three conclusions. First, a lucky accident can constitute a legitimate explanation for the success of a scientific theory. This challenges the realist view, which typically treats coincidences as mere rhetorical devices aimed at discrediting truth-independent accounts of empirical success, while also adding to antirealist perspectives, which have largely neglected the explanatory role of coincidences in science.\footnote{Although Drude's theory is a striking example of success achieved through error cancellation, additional cases can be found even when limiting the discussion to electrons in metals. For example, in the density-functional theory predictions of the structure and energetics of metal surfaces, the local-density approximation (LDA) to the exchange-correlation density functional is systematically more successful than, e.g., the gradient-corrected Perdew-Burke-Ernzerhof (PBE) approximation, despite PBE being closer to the exact (true) exchange-correlation functional than LDA. As noted by \cite{Patra2017},  ``LDA displays a remarkable error cancellation between its exchange and correlation contributions. Usually, the LDA exchange energy contribution to the surface energy is an overestimate, while the correlation contribution is a significant underestimate, and their combination results in an accurate prediction. PBE improves both the exchange and the correlation contributions, but loses the remarkable error cancellation of LDA.'' (p.\ 9190).} Second, the occurrence of success-by-coincidence stands in contrast to success-by-truth, insofar as increasing the truth content of a theory does not necessarily enhance its empirical success. This has been demonstrated by incorporating `true' elements---such as the correct electron velocity or heat capacity---into the otherwise false Drude's theory, resulting in a quantitative and qualitative deterioration of its empirical adequacy. Third, this coincidental explanation can operate in tandem---not in opposition---with other antirealist explanations, offering a deeper account of scientific success. This was illustrated by the interplay between success-by-coincidence and success-by-predictive similarity: while the predictions of Drude's theory are close to those of Sommerfeld's theory, it is a coincidence that explains their similarity.

To conclude, it is important to emphasize that I do not claim that coincidences provide a universal explanation for scientific success. Likewise, I do not deny that some scientific theories may enjoy success by virtue of their approximate truth. What this case study illustrates, however, is that success can arise through multiple pathways that do not depend on truth---including some that may appear surprising---suggesting that the ability of a mature theory to achieve predictive success may not always be sufficient to distinguish approximately true theories from false ones.

\newpage
%\section{References}
\bibliographystyle{apalike}
\bibliography{Final-Bibliography}
\end{document}